\documentclass[aps,pra,showpacs,twocolumn]{revtex4}
\usepackage{graphicx}

\begin{document}

\title{A high-flux source of polarization-entangled photons from a periodically-poled KTP parametric downconverter}

\author{Christopher E. Kuklewicz, Marco Fiorentino, Ga\'{e}tan Messin,\\
Franco N. C. Wong, and Jeffrey H. Shapiro}

\affiliation{Research Laboratory of Electronics, Massachusetts Institute of Technology, Cambridge, Massachusetts 02139}

\begin{abstract}
  We have demonstrated a high-flux source of polarization-entangled
  photons using a type-II phase-matched periodically-poled KTP
  parametric downconverter in a collinearly propagating configuration.
  We have observed quantum interference between the single-beam
  downconverted photons with a visibility of 99\% and a measured
  coincidence flux of 300/s/mW of pump.  The
  Clauser-Horne-Shimony-Holt version of Bell's inequality was violated
  with a value of $2.711 \pm 0.017$.
\end{abstract}

\pacs{42.65.Lm, 03.65.Ud, 42.50.Dv}

\maketitle

\section{Introduction} 
\label{sec:introduction}

Entanglement is the basis for fundamental demonstrations of quantum
mechanics such as violation of Bell's inequality and tests of local
realism \cite{Bell,CHSH}.  Moreover, it is essential to a wide variety
of quantum communication applications, including teleportation
\cite{Bennett,ShapiroNJP} and quantum secret sharing \cite{Hillery},
and it can be used for quantum key distribution \cite{Ekert}.  A
high-flux source of polarization-entangled photons is therefore
desirable for practical implementation of a variety of
entanglement-based applications.  Spontaneous parametric
downconversion (SPDC) in a noncollinearly propagating, angle
phase-matched crystal, such as beta barium borate (BBO), is often used
to generate polarization entanglement
\cite{Kwiat99,Weinfurter,Kwiat95}.  However, only a small segment of
the output cone of the downconverted photons is collectible.
Moreover, angle phase matching precludes the use of long crystals for
more efficient generation or for narrowband generation in a cavity
configuration.

We take a different, yet simple, approach to entanglement generation
that is based on periodically-poled potassium titanyl phosphate
(PPKTP) with collinear propagation of the pump, signal, and idler
fields.  A periodically-poled nonlinear crystal such as PPKTP, with an
appropriate grating period, permits efficient three-wave mixing at
user-selectable wavelengths within the crystal's transparency window
by the technique of quasi-phase matching (QPM).  Under QPM one can
choose to propagate along a principal axis of the crystal, thus
avoiding undesirable angle walkoff and permitting collinear
propagation in long crystals, which can be utilized in cavity
configurations for enhancing downconversion efficiency and providing
high-brightness narrowband outputs \cite{ShapiroWong}.  Furthermore, a
single-beam configuration of co-propagating signal and idler photons
simplifies the transport of entangled photon pairs.  For the current
work we report measurements of single-beam quantum interference with a
visibility of up to 99\% and Bell's inequality violation from a
continuous-wave (cw) PPKTP parametric downconverter.

\section{Single-beam entanglement} 
\label{sec:single-beam}

Figure~1 shows the basic concept of our cw collinearly propagating
single-beam SPDC for polarization entanglement.  It consists of a
length-$L$ type-II phase-matched PPKTP crystal followed by a
length-$L/2$ KTP timing compensator.  Consider the probability
amplitudes of a single pair of downconverted signal and idler photons
that originate, via quantum superposition, from two locations, $A$ and
$B$, which are symmetrically displaced from the center of the
periodically-poled crystal.  Because of the crystal's birefringence,
the horizontally ($H$) polarized signal photon along the crystal's $y$
axis always exits the crystal ahead of the conjugate idler photon,
which is vertically ($V$) polarized along the crystal's $z$ axis.  The
timing compensator in Fig.~1 is a KTP crystal that is oriented similar
to the PPKTP crystal but with the $y$ and $z$ axes interchanged.
Time-resolved signal and idler photon counting before the compensating
crystal will collapse the quantum superposition.  This will tag the
location---$A$ or $B$---at which the pair was created by virtue of the
time difference between the signal and idler counts.  After the
compensator, the signal and idler pair is at a quantum superposition
of locations $C$ and $D$, with either $H_A$ leading $V_A$ or $V_B$
leading $H_B$ by the same time interval.  The temporal information has
been erased because the separation of the photon pair no longer
reveals the source location in the PPKTP\@.  Although the timing
compensator has erased one feature that distinguishes the signal and
idler photons, they are still distinguishable by virtue of their
orthogonal polarizations.  This polarization information can be
erased, however, by rotating the output polarizations by $\pi/4$,
before analysis along $H$ and $V$.  The erasure of such identifying
information is essential to all quantum interference experiments
\cite{Rubin}.  Furthermore, it is also necessary to erase any
frequency or spatial mode information, which might distinguish between
photons in a detected pair, by collecting only collinear and
degenerate pairs.  Ignoring the vacuum and higher photon-number
components, the state at the output of this polarization rotation is
the biphoton,
\begin{equation}
|\psi\rangle = (|H\rangle_1 |H\rangle_2 - |V\rangle_1 |V\rangle_2)/\sqrt{2}\,.
\end{equation}
The photons labeled 1 and 2 can be analyzed with a polarization beam
splitter (PBS) and a null should occur in coincidence measurements.
This coincidence null is equivalent to the coincidence dip in
Hong-Ou-Mandel (HOM) interferometry \cite{Hong,Rubin,Kuklewicz}.

It is possible to utilize the single-beam SPDC output of Fig.~1 to
obtain polarization-entangled photons with a 50/50 beam splitter.
Each photon of the orthogonally-polarized photon pair that is
generated from the PPKTP is equally likely to be transmitted or
reflected at the beam splitter.  Hence half of the generated pairs
yield one photon in the transmitted path and one in the reflected
path.  If we post-select only these events, the two photons in the two
paths are in a polarization-entangled triplet state:
\begin{equation}
|\psi\rangle = (|H\rangle_T |V\rangle_R + |V\rangle_T |H\rangle_R)/\sqrt{2}\,,
\end{equation}
where the subscripts $T$ and $R$ refer to the transmitted and
reflected paths of the 50/50 beam splitter, respectively.  We note
that post selection is not necessary if quantum memories that allow
non-destructive loading verification are used \cite{Lloyd}.

\section{Single-beam quantum interference} 
\label{sec:interference}

We have implemented the single-beam downconversion scheme in the setup
shown in Fig.~2.  A 10-mm-long flux-grown PPKTP crystal with a grating
period of 8.84 $\mu$m was used for frequency-degenerate type-II
quasi-phase matched operation at a pump wavelength of 397 nm.  The
output wavelength at 795 nm was chosen to match the transition
wavelength of the $D_1$ line of Rb, which has been proposed for use as
a trapped-atom quantum memory for long-distance teleportation
\cite{Lloyd}.  The 1-mm-thick PPKTP crystal was anti-reflection coated
at 397 and 795 nm and was pumped with a 10-mW cw external-cavity
ultraviolet (UV) diode laser that was weakly focused to a
$\sim$200-$\mu$m beam waist in the crystal.  The crystal was set up
for collinear propagation along its $x$ axis and with its $y$ axis
aligned with the pump's horizontal polarization.  For type-II phase
matching, the nonlinear coefficient $d_{24}$ was utilized, and the
signal and idler outputs were orthogonally polarized along the
crystal's $y$ and $z$ axes, respectively.  The operating temperature
of the PPKTP crystal was controlled with a thermoelectric cooler that
allowed us to tune the crystal to exact frequency degeneracy and was
typically set at 20$^{\circ}$C\@.  After passing through the 5-mm-long
KTP timing compensator to erase the timing information, the output
beam was sent through an interference filter centered at 795 nm that
had a 1-nm bandwidth and 80\% in-band transmission.  We have imaged
the downconverted light through the 1-nm interference filter onto a
high-sensitivity CCD camera, and found this light had a divergence
full angle of $\sim$20 mrad, in good agreement with theoretical
estimates of the external divergence angle for our PPKTP system.  Due
to our propagation along one of the principal axes in PPKTP, the
observed divergence angle of 20-mrad/nm of bandwidth is more than an
order of magnitude larger than that for the usual angle phase-matched
configuration in BBO for a given crystal length and spectral bandwidth
\cite{Weinfurter}.

There were some UV-induced fluorescence photons from the flux-grown
PPKTP crystal.  These fluorescence photons, estimated to be
$\sim$1000/s/mW of pump generated in a 1-nm bandwidth and collected
through a small iris, increased the singles rates by $\sim$5\%.
Accidental coincidences caused by them were relatively insignificant
in our experiments.  We installed two dichroic mirrors that passed the
795-nm outputs but attenuated the UV pump by $\sim$40 dB to minimize
additional fluorescence from the KTP timing compensator and other
optical elements.  An adjustable iris was used to control the
effective divergence angle of the transmitted beam.  A smaller iris
increased the depth of field and reduced the spatial resolution of the
output photons such that the photons generated from locations $A$ and
$B$ in Fig.~1 would be spatially indistinguishable.  The iris also
served to block off-axis, nearly-degenerate photon pairs that would
not contribute to either quantum interference or polarization
entanglement.

For quantum-interference measurements, the 50/50 beam splitter in
Fig.~2 was not necessary.  Removing this beam splitter improves the
conditional detection probability by a factor of two, and we have made
measurements with and without it present.  With the 50/50 beam
splitter present, each output beam was sent through a half-wave plate
(HWP), a PBS for polarization analysis, and a clean-up polarizer to
eliminate the leakage of horizontally-polarized light into the
vertically-polarized path.  Also, prisms were used in the
horizontally-polarized output paths to reduce accidental counts due to
the UV pump.  All four PBS outputs were focused on commercial Si
single-photon counting detectors, whose detection quantum efficiencies
are estimated to be 50-55\% at 795 nm and whose dark count rates are
less than 100/s.  We then combined the $\sim$35-ns-long electrical
pulses from detectors $TT$ and $TR$ with AND logic, and both the
singles rates from the detector output pulses and the coincidence rate
from the logic pulses were counted.  The dead times of our
single-photon counters (including the pulse width) were measured to be
$\sim$50 ns, which is negligible at our operating count rates of
10$^5$/s or less.  By blocking the pump and measuring the singles and
coincidence rates due to different amounts of stray light for the four
detectors, we were able to determine the effective coincidence window
for each pair of detectors and calibrate the accidental coincidence
rates, for removal in post-detection data analysis.  Typical rates,
with the 50/50 beam splitter in place, were $12,000$ singles/s, pair
coincidences of 1,200/s in a $\sim$70-ns coincidence window, and
accidental coincidences of $\sim$12/s for 10 mW of pump and 1-nm
detection bandwidth.  The main contribution to the accidental
coincidences is the Poisson occurrence of a double pair within the
coincidence window in which the signal of one pair and the idler of
the other pair are detected.  The reliability of the logical AND
coincidence detection and the data analysis methodology was checked by
comparing it to a high time-resolution (sub-ns) start-stop histogram
obtained with a Picoquant TimeHarp 200.

Consider the case without the 50/50 beam splitter and with only the
detectors labeled $TT$ and $TR$ in Fig.~2 in use.  PBS$_T$ in the
transmitted path and the crystal's $y$ and $z$ axes were aligned such
that if the HWP$_T$'s fast and slow axes were also aligned the same
way, the signal and idler photons would be separately detected to
yield the pair generation rate.  When HWP$_T$ was set at $\theta_T =
0$ to yield zero polarization rotation and the iris was open, we
observed a coincidence rate as high as $46,100$/s, from which we infer
a pair generation rate of $\sim$10$^6$/s using our 21\% conditional
detection efficiency. At $\theta_T = \pi/8$, the output beam underwent
a $\pi/4$ polarization rotation, and each incident photon had a 50/50
chance of being transmitted or reflected at PBS$_T$.  For photon pairs
that were spectrally, spatially, and temporally indistinguishable, the
state after HWP$_T$ was given by Eq.~(1), and quantum interference
between the signal and idler of a photon pair occurred, resulting in a
reduction in the coincidence rate.  When we made the photons
distinguishable, by frequency detuning, no quantum interference
occurred and a coincidence rate of 50\% relative to the zero-rotation
rate was observed.  With the iris open and $\theta_T=\pi/8$, we
observed a coincidence rate of 11,800/s, corresponding to a visibility
$V = [C(0)-C(\pi/8)]/[C(0)+C(\pi/8)]$ of 59\%, where $C(\theta_T)$ is
the coincidence rate at a HWP$_T$ angle setting of $\theta_T$. As we
reduced the size of the iris, the effective divergence angle was
reduced and the depth of field improved, leading to a reduction in the
coincidence rates and an increase in the visibility.  We have achieved
$V=97.7\%$ with a 200-$\mu$m-diameter aperture, corresponding to a
divergence full angle of $\sim$2 mrad, for the flux-grown PPKTP
downconverter.

We have also made similar quantum interference measurements with a
second PPKTP crystal, which was hydrothermally grown and had a grating
period of 9.01 $\mu$m.  This 10-mm-long crystal was pumped with the
second harmonic of a Ti:sapphire laser centered at 397 nm with a
maximum usable power of $\sim$30 mW\@.  The tunable UV pump source and
the temperature tuning of the PPKTP crystal permitted us to maintain
the SPDC operating point at exact frequency degeneracy for the
collinearly propagating portion of the output.  Typical pump powers
were 5 mW and the crystal temperature was usually set at
30$^{\circ}$C\@.  We added a collimating lens for the output beam and
an adjustable iris was used to control the depth of field.  The
conditional detection efficiency for the hydrothermally grown PPKTP
setup was $\sim$25\% for a 3-mm-diameter aperture and a 3-nm
interference filter.  Figure 3 shows the quantum-interference
visibility as a function of the aperture size for four different
interference-filter bandwidths.  At the aperture size of 1 mm, which
is equivalent to a divergence full angle of 5.4 mrad, we observed a
visibility of $99\pm 1$\% for the 1-nm filter.  Figure 3 clearly shows
that for larger aperture sizes, which correspond to shallower depths
of field at the crystal, the visibility is reduced.  This reduction
occurred because spatial and spectral indistinguishability was no
longer fully maintained.  The inset in Fig.~3 shows the detected
coincidence rate as a function of the aperture size for the case of
the 1-nm filter.  At the highest visibility level of 99\%, obtained
with a 1-mm aperture, the measured coincidence rate was $\sim$300/s/mW
of pump power, which is one of the highest reported values at
near-unity visibility level for an entanglement source
\cite{Kwiat99,Weinfurter}.  Moreover, for an aperture size of 3 mm,
with a corresponding divergence angle of 16 mrad, we measured a flux
of over 5000/s/mW of pump while maintaining a visibility of 90\%.  We
should note that the hydrothermally grown PPKTP crystal was found to
be more efficient (with $d_{\rm eff}\sim 1.60$ pm/V) than the
flux-grown PPKTP crystal ($d_{\rm eff}\sim 0.74$ pm/V).  In addition,
the UV-induced fluorescence of the hydrothermally grown PPKTP was
about 25\% of that for the flux-grown crystal.

\section{Polarization entanglement} 
\label{sec:entanglement}

One can easily obtain polarization-entangled photon pairs using the
experimental setup in Fig.~2.  Each member of an
orthogonally-polarized photon pair that is generated in the PPKTP
downconverter has a 50\% chance of being transmitted or reflected by
the 50/50 beam splitter.  When one photon appears in the transmitted
path and one in the reflected path---something that can be
post-selected by monitoring for coincidences---the joint state of the
beam splitter's output is the polarization-entangled triplet given by
Eq.~(2).

The quality of the single-beam polarization entanglement can be
evaluated by measuring the violation of the Clauser, Horne, Shimony,
and Holt (CHSH) form of Bell's inequality \cite{CHSH}.  We have made
such Bell's inequality measurements using the diode-pumped flux-grown
PPKTP downconverter with a 1-nm interference filter and $\sim$10 mW of
pump power.  Referring to Fig.~2, the light in the transmitted path
and in the reflected path of the 50/50 beam splitter were separately
analyzed with a HWP and a PBS\@.  Simultaneous coincidence
measurements between detectors $TT$ and $RT$ ($C_{TT,RT}$), detectors
$TR$ and $RR$ ($C_{TR,RR}$), and detectors $TT$ and $TR$
($C_{TT,TR}$), as indicated in Fig.~2, were taken for two different
$\theta_T$ settings of 0 and $\pi/8$ for the transmitted beam.  The
HWP$_T$'s angle setting $\theta_T$ was ascertained by
quantum-interference measurements in the transmitted path using
$C_{TT,TR}$.  At each $\theta_T$ angle, coincidence measurements
$C_{TT,RT}$ and $C_{TR,RR}$ over a 10-s interval were taken at 32
different positions of the HWP$_R$'s setting ($\theta_R$ at
$\sim$$\pi/16$ intervals) in the reflected path of the 50/50 beam
splitter.  These coincidence measurements were used to calculate the
value of the CHSH inequality.  Figure~4 shows the coincidence counts
$C_{TT,RT}$ for $\theta_T = 0$ and for $\theta_T =\pi/8$ and their
sinusoidal fits, showing visibilities of 98\% and 93\%, respectively.

The four coincidence-count data sets of $C_{TT,RT}$ and $C_{TR,RR}$
for $\theta_T = 0$ and $\theta_T = \pi/8$ are fit to sinusoidal
functions.  Using these four fits (with their estimated errors) we
construct the CHSH expectation $E$ functions and the $S$ parameter
function \cite{Kwiat99,Kwiat95}.  $E$ is defined by:
\begin{widetext}
\begin{equation}
  \label{eq:E}
E(\theta_T,\theta_R)=\frac
{C_{TT,RT}(\theta_T,\theta_R) + C_{TR,RR}(\theta_T,\theta_R)-   C_{TT,RR}(\theta_T,\theta_R) - C_{TR,RT}(\theta_T,\theta_R)}
{C_{TT,RT}(\theta_T,\theta_R) + C_{TR,RR}(\theta_T,\theta_R) + C_{TT,RR}(\theta_T,\theta_R) + C_{TR,RT}(\theta_T,\theta_R)}\,.
\end{equation}
\end{widetext}
Note that $E$ depends on $C_{TT,RR}$ and $C_{TR,RT}$ which were not
directly measured, so we derive their values from the fits of
$C_{TT,RT}$ and $C_{TR,RR}$ with $\theta_R \rightarrow
\theta_R+\pi/4$:
\begin{eqnarray}
C_{TT,RR}(\theta_T,\theta_R) & = &
C_{TT,RT}\left(\theta_T,\theta_R+\frac{\pi}{4}\right)\,, \\
C_{TR,RT}(\theta_T,\theta_R) & = &
C_{TR,RR}\left(\theta_T,\theta_R+\frac{\pi}{4}\right)\,.
\end{eqnarray}
The parameter $S$ is composed of $E$ functions for two values of
$\theta_R$ and two values for $\theta_T$.  In our case $\theta_T = 0,\
\pi/8$ and $\theta_R = \pi/16,\ 3\pi/16$. Thus our $S$ parameter is
defined as:
\begin{equation}
  \label{eq:S}
S = \left| E\left(0,\frac{\pi}{16}\right) -
  E\left(\frac{\pi}{8},\frac{\pi}{16}\right) + E\left(0,\frac{3\pi}{16}\right) +
  E\left(\frac{\pi}{8},\frac{3\pi}{16}\right) \right |\,.
\end{equation}
Classical and hidden variable theories predict $S\leq 2$, while
quantum mechanics permits $S \leq 2\sqrt{2} \approx 2.828$, with
equality occurring for maximal polarization entanglement, i.e., a
polarization Bell state such as the triplet from Eq.~(2).  We obtained
an $S$ value of $2.711 \pm 0.017$, which indicates good polarization
entanglement of our PPKTP SPDC source.

\section{Conclusion} 
\label{sec:conclusion}

In summary, we have demonstrated a cw high-flux source of
polarization-entangled photons using a PPKTP parametric downconverter
in a collinearly propagating configuration.  The single-beam output is
shown to allow easy control of its spatial and spectral contents and
simplify the transport of the photon pairs.  By using a circular
aperture we were able to obtain a high visibility of 99\% with a
corresponding flux of $\sim$300/s/mW of pump power.  Polarization
entanglement was obtained from the cw single-beam PPKTP downconverter
with a 50/50 beam splitter, and we measured Bell's inequality
violation with $S = 2.711 \pm 0.017$.  We have found that the use of
periodically-poled nonlinear material and the single-beam collinearly
propagating configuration offer distinct advantages over the usual
noncollinearly phase-matched BBO downconverter.  Wavelength tunability
of the paired photons with no change in the output beam angle can be
easily accomplished with temperature tuning of the crystal and/or a
change in the pump wavelength.  Long crystals with collinear outputs
can be used to allow more efficient generation and collection of
entangled photons.  We expect that future entanglement sources based
on cavity-enhanced parametric downconversion in long crystals can
significantly improve on its flux and also its spatial and spectral
contents \cite{ShapiroWong}.

\section{ Acknowledgments} 
\label{sec:acknowledgments}

This work was supported by the DoD Multidisciplinary University
Research Initiative (MURI) program administered by the Army Research
Office under Grant DAAD-19-00-1-0177, and by the National
Reconnaissance Office.

\newpage\vspace*{2in}\begin{figure}[h]
\centerline{\includegraphics[width=\hsize]{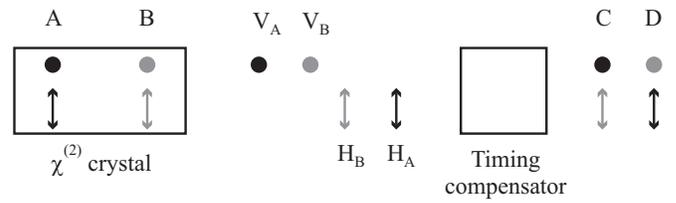}}
\caption{Schematic for generating entanglement in a single optical
  beam.  A single signal/idler photon pair is generated, via quantum
  superposition, at locations $A$ and $B$.  The timing of the
  probability amplitudes of the orthogonally-polarized outputs from
  locations $A$ and $B$ in a $\chi^{(2)}$ crystal is shown before and
  after a timing compensation crystal.  Timing information is erased
  for photon pairs at locations $C$ and $D$. Horizontal (vertical)
  polarization: $\updownarrow$ ($\bullet$).}
\end{figure}

\newpage\vspace*{2in}\begin{figure}[h]
\centerline{\includegraphics[width=\hsize]{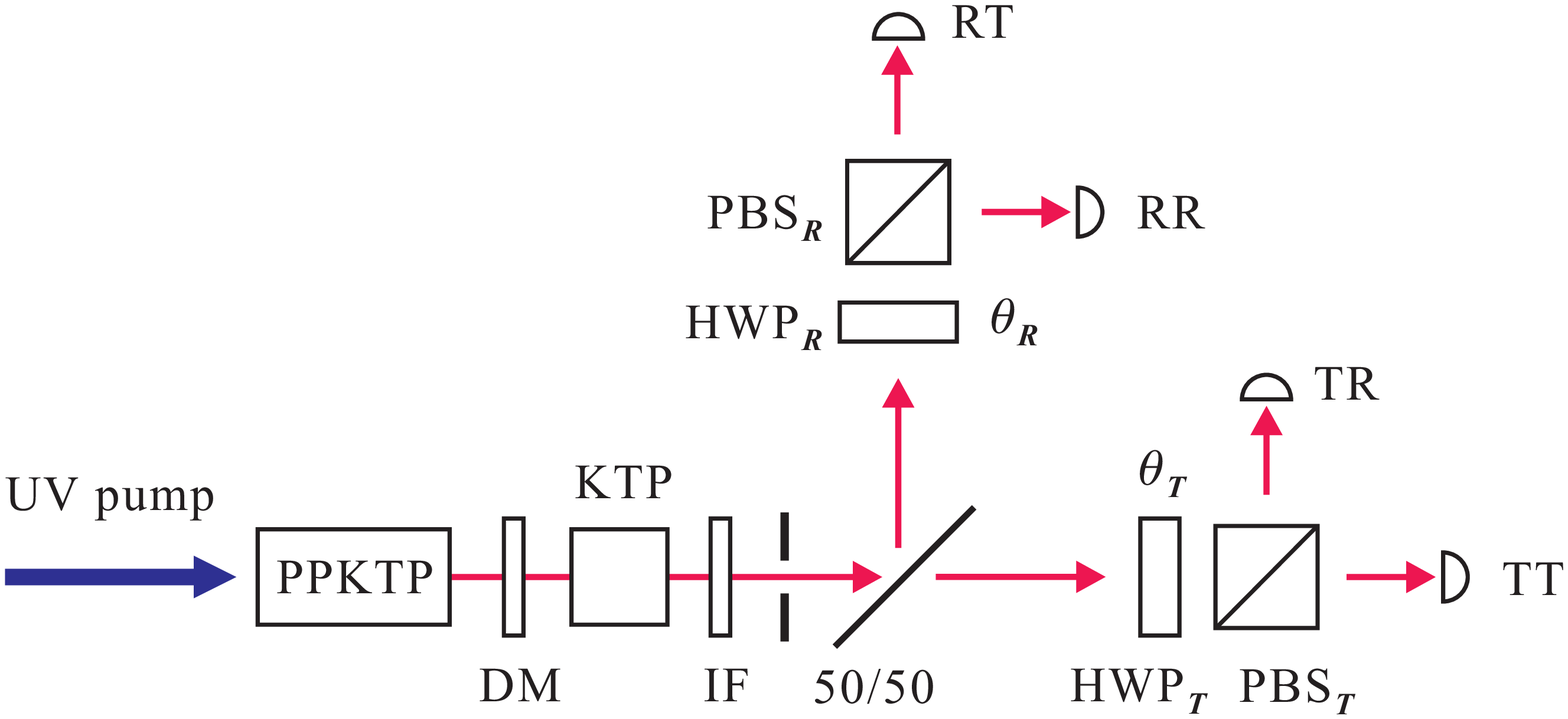}}
\caption{Schematic of experimental setup.
  The 50/50 beam splitter is removed for quantum interference
  measurements which are made with detectors $TT$ and $TR$. Detectors
  $TT$ and $RT$ are used for Bell's inequality measurements.  DM =
  dichroic mirror; IF = interference filter; HWP = half-wave plate;
  PBS = polarizing beam splitter.}
\end{figure}

\newpage\vspace*{2in}\begin{figure}[h]
\centerline{\includegraphics[width=\hsize]{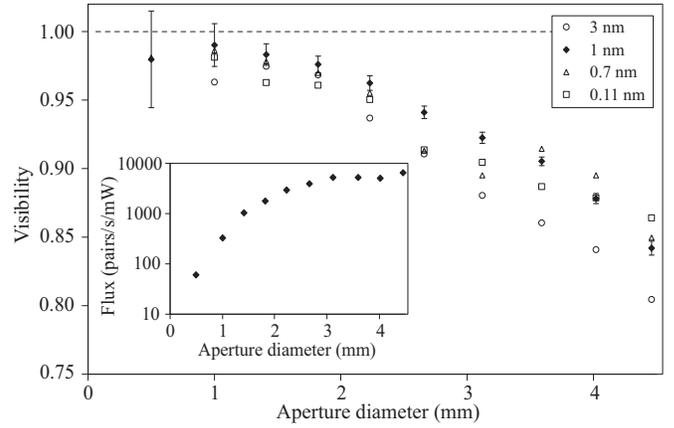}}
\caption{Plot of visibility $V$ as a function
  of aperture diameter for interference filter bandwidths of 3, 1,
  0.7, and 0.11 nm.  The 0.7-nm filter was composed of two identical
  1-nm filters in series.  Error bars for the 1-nm data are displayed.
  Inset plots the detected coincidence flux as a function of the
  aperture size for the filter bandwidth of 1 nm, showing the
  trade-off between usable flux and visibility.}
\end{figure}

\newpage\vspace*{2in}\begin{figure}[h]
\centerline{\includegraphics[width=\hsize]{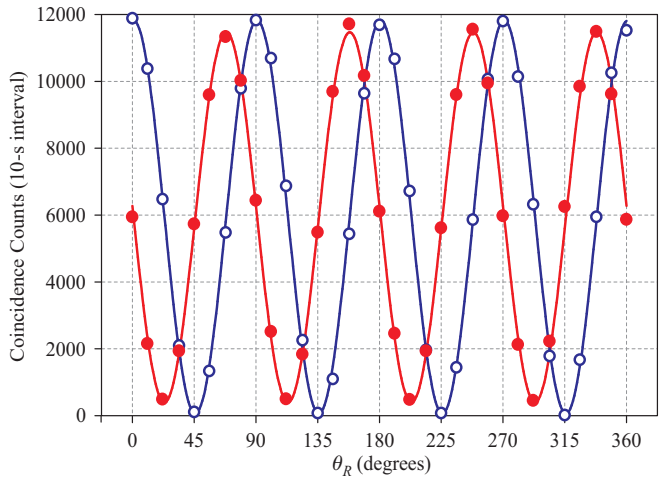}}
\caption{Plot of coincidence counts for $\theta_T = 0$ (open
  circles) and $\theta_T = \pi/8$ (solid circles) in a 10-s counting
  interval as a function of HWP$_R$ setting $\theta_R$ in the
  reflected path of the 50/50 beam splitter in Fig.~2.  Accidental
  coincidences have been removed in these plots and the sinusoidal
  fits (solid lines) are used for obtaining visibility and the $S$
  parameter in the CHSH inequality.}
\end{figure}
\end{document}